\documentclass[12pt]{article}
\usepackage{graphicx}
\usepackage{booktabs}
\usepackage{amsmath,amssymb,amsthm,graphicx,url}

\usepackage{times}
\usepackage{pdflscape}

\newcommand{\pandc}{packing and cracking}
\newcommand{\spc}{SPC}

\newcommand{\bp}{\boldsymbol{p}}
\newcommand{\bl}{\boldsymbol{\ell}}
\newcommand{\op}{p^\circ}
\newcommand{\bop}{\boldsymbol{p}^\circ}
\newcommand{\ol}{\ell^\circ}
\newcommand{\bol}{\boldsymbol{\ell}^\circ}
\newcommand{\np}{p^*}

\newcommand{\nl}{\ell^*}
\newcommand{\bnl}{\boldsymbol{\ell}^*}

\theoremstyle{definition}
\newtheorem{theorem}{Theorem}

\newtheorem{example}[theorem]{Example}

\newtheorem{remark}[theorem]{Remark}

\makeatletter
\g@addto@macro\bfseries{\boldmath}
\makeatother

\title{Gerrymandering and the net number of US House seats won due to
  vote-distribution asymmetries}

\author
{Jeffrey S. Buzas and Gregory S. Warrington$^{\ast}$\\
\\
\normalsize{Department of Mathematics \& Statistics, University of Vermont,}\\
\normalsize{16 Colchester Ave., Burlington, VT 05401, USA}\\
\\
\normalsize{$^\ast$To whom correspondence should be addressed; E-mail:  gswarrin@uvm.edu.}
}

\date{}

\begin{document} 



\maketitle 


\begin{abstract}
Using the recently introduced declination function, we estimate the
net number of seats won in the US House of Representatives due
to asymmetries in vote distributions. Such asymmetries can arise from 
combinations of partisan gerrymandering and inherent geographic
advantage. Our estimates show significant biases in favor of the
Democrats prior to the mid 1990s and significant biases in favor of
Republicans since then. We find net differences of 28, 20 and 25 seats
in favor of the Republicans in the years 2012, 2014 and 2016,
respectively. The validity of our results is supported by the
technique of simulated packing and cracking. We also use this
technique to show that the presidential-vote logistic regression model
is insensitive to the packing and cracking by which partisan
gerrymanders are achieved.
\end{abstract}


\section{Introduction}

The partisan composition of the US House of Representatives is the
result of a number of factors: the economy, social issues, who is
president, party platforms, voter ID laws, propaganda, campaign
finance laws, the characteristics and incumbency statuses of
individual candidates, and district plans, to name
a few. Our focus in this paper is on district plans.

A partisan gerrymander is, by definition, a district plan that enables
a party to win more seats than it would have under a ``neutral''
district plan. It is sometimes easy to detect a partisan
gerrymander. For example, a North Carolina legislator openly admitted
that his committee planned to redraw districts with a goal of partisan
advantage~\cite{lewis}. Other times the convoluted shapes of districts
provide evidence that is almost as compelling. In fact, significant
energy has been devoted to developing geometric methods of identifying
gerrymandered districts (see~\cite{compactness} for an overview). But
none of these geometric approaches provide a direct way to determine
the number of extra seats that have been won through the gerrymander
--- the fundamental purpose of a partisan gerrymander.

In the first part of this paper we use the declination function,
$\delta$, introduced by the second author in~\cite{declination}, to
estimate the effect of asymmetries in how district plans treat the two
major parties on the number of House seats won by each. To do so, we
explore more thoroughly the scaled function $5N\delta/12$ which we
refer to in this paper as the $S$-declination. (The variable, $N$,
denotes the number of districts in the election; ``$S$'' is intended
to remind the reader that this scaling counts \emph{seats}.) We argue
that the $S$-declination provides a good estimate of the number of
extra seats won through these partisan asymmetries.

\begin{remark}
  In~\cite{declination} we consider a similar scaling of
  $N\delta/2$. While simpler, this appears to slightly overcount the
  number of seats in a given state and year. The effect is not
  significant when single-state US House elections are considered as
  is done in~\cite{declination}, but in the typically larger state
  legislatures or when considering national effects, the difference
  between the two scalars is non-trivial.
\end{remark}

Table~\ref{tab:net} presents the results of applying the
$S$-declination to the US House elections since 1972. The data
indicate that, on a national level, while the Democrats consistently
benefited through the early 1990s from partisan asymmetry in district
plans, the situation has been reversed since the late 1990s.


There have been a number of previous attempts to estimate the net
effect of gerrymandering on a national level (see, e.g., the
references in~\cite{cottrell}). The most recent of which we are aware
is a report~\cite{brennan} by Royden and Li that considers House
elections since 2012. Our estimate for the year 2012 of 28 extra
republican seats falls within 25--36, the narrower of the two ranges
provided by Royden and Li.  However, our estimate differs markedly
from the estimate of approximately one seat arrived at in a recent
paper~\cite{cottrell} by Chen and Cottrell.  (See
also,~\cite[pgs. 36,37]{Wang} whose results are intermediate to those
of the mentioned studies.)

In the second part of this paper we attempt to explain the discrepancy
between the Chen-Cottrell estimate and the two other estimates. Our
main tool will be a simulated packing and cracking technique that we
introduce in Section~\ref{sec:background}. We first use this technique
to validate the $S$-declination as a means of counting extra seats.
We then use it to show that the logistic-regression approach of Chen
and Cottrell is insensitive to the main technique of partisan
gerrymandering.

The structure of this paper is as follows. In
Section~\ref{sec:background} we introduce the necessary background and
terminology on both partisan gerrymandering as well as on packing and
cracking. We also describe a simple framework for simulated packing
and cracking that we use in later sections to gauge how faithfully
various methods register partisan gerrymandering efforts. In
Section~\ref{sec:declination} we review the declination function and
show using the packing and cracking technique that the declination
faithfully registers partisan gerrymandering. In Table~\ref{tab:net}
we record the net number of seats in House elections since 1972 that,
according to the declination, should be attributed to partisan
asymmetry in the vote distribution. In Section~\ref{sec:recent} we
show that the model utilized in~\cite{cottrell} is insensitive to
\pandc, thereby explaining the discrepancy between their estimate and
the estimates of both this paper and~\cite{brennan}. We end with our
conclusion.

\section{Background}
\label{sec:background}

Partisan gerrymanders have historically been recognized by identifying
individual districts with contorted boundaries. The primary advantage
of this approach is that wild shapes are immediately convincing as
evidence of nefariousness. Unfortunately, there are several
disadvantages. First, there are many valid reasons for a district to
have a strange shape (e.g., geographic constraints, the Voting
Rights Act, and county boundaries). Second, gerrymandering can occur
without particularly unusual boundaries. Third, the shapes of the
districts do not directly tell one anything about how effective the
gerrymander is in winning additional seats for one party. An
alternative approach to the study of district shapes is to analyze the
distribution of votes among the various districts. In this section we
introduce the notation and terminology necessary to work with these
distributions and to see the effect partisan gerrymandering has on
them. We assume that each district has the same number of total votes.

\begin{figure}
  \centering
  \includegraphics[width=.8\linewidth]{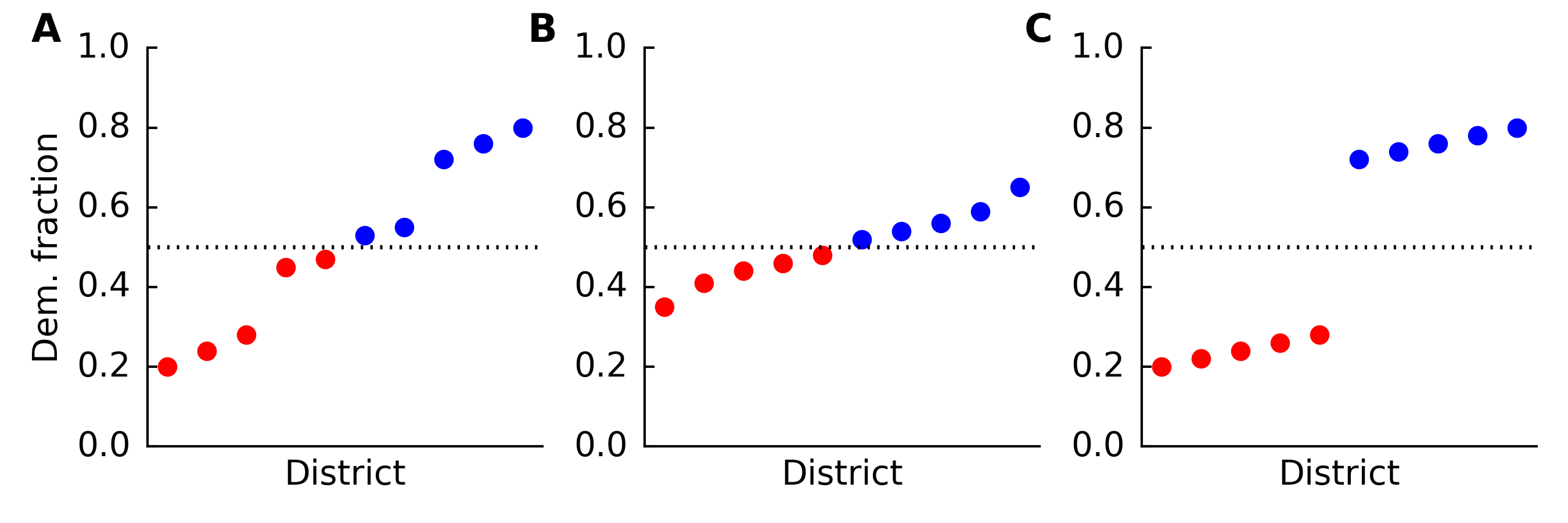}
  \caption{Symmetric vote distributions for fair elections
    with equally popular parties.}
  \label{fig:plot}
\end{figure}

We define an \emph{$N$-district election} to be a weakly increasing
sequence $\bl = (\ell_1,\ell_2,\ldots,\ell_N)$ in which $\ell_i$
indicates the democratic fraction of the two-party (legislative) vote
in district $i$. We visualize $\bl$ by plotting a point $v_i =
(i/N-1/2N,\ell_i)$ for each $i$. Figure~\ref{fig:plot} illustrates
plots of $\bl$ for three hypothetical elections.

\begin{figure}
  \centering
  \includegraphics[width=.8\linewidth]{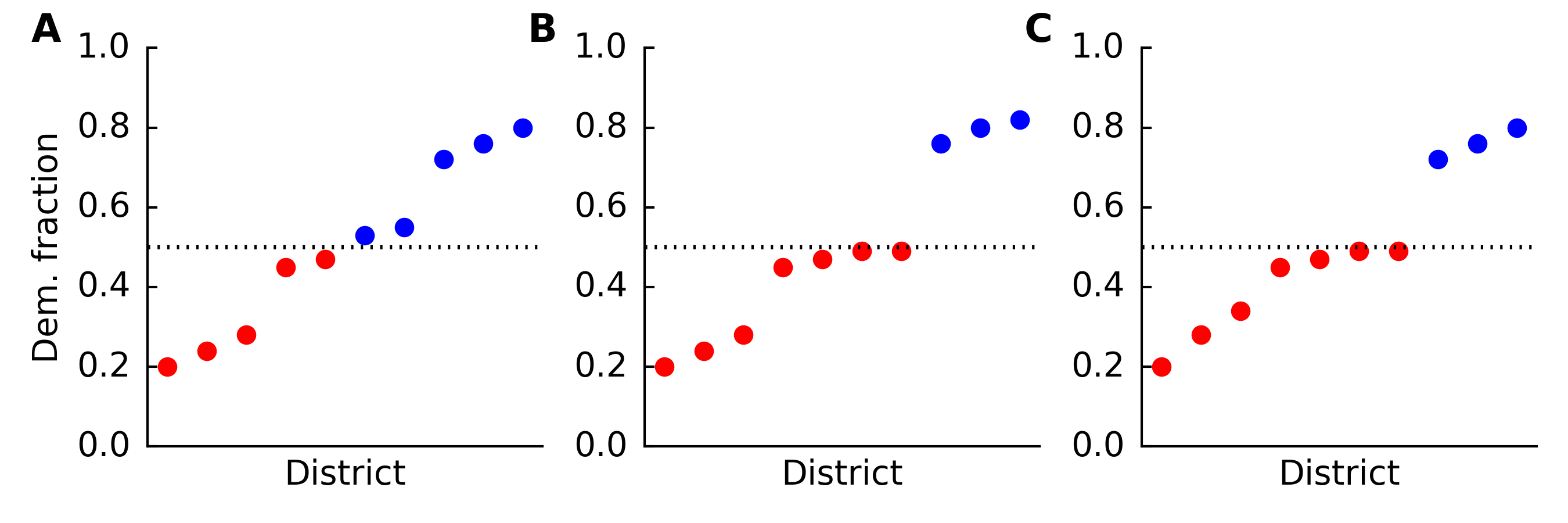}
  \caption{{\bf (A)} Repeat of election from Figure~1.A. {\bf (B)}
    Illustration of packing of the distribution from (A) for a gain of
    two republican seats.  {\bf (C)} Illustration of cracking of the
    distribution from (A) for a gain of two republican seats. }
  \label{fig:pandc}
\end{figure}

While researchers are still working on robust tests both for
identifying gerrymanders and for assessing the effects of
gerrymandering, \emph{how} partisan gerrymandering occurs is well
understood: Parties create partisan gerrymanders by ``packing and
cracking'' votes. Suppose the Democrats are in control of
redistricting and the Republicans are poised to win district $X$. In
\emph{packing}, Republicans are moved from $X$ to other districts in
which the Republicans already have enough strength to win. These votes
are effectively wasted in the new districts while district $X$ falls
to the Democrats. \emph{Cracking} works similarly, except now the
Republicans are spread among districts that they have no chance of
winning. Once the cracking occurs, the recipient districts are lost by
the Republicans by smaller margins, but they are still
lost. Figure~\ref{fig:pandc} illustrates the effect of \pandc\ on the
vote distribution illustrated in Figure~\ref{fig:plot}.A.

\subsection{Simulated packing and cracking}
\label{subsec:mpandc}

In order to validate the $S$-declination (see
Section~\ref{sec:declination}) as a measure of the number of extra
seats won under a partisan gerrymander, we will examine how
perturbations of a vote distribution by packing and/or cracking affect
the value of the $S$-declination. The simple technique we will use is
that of \emph{simulated packing and cracking (\spc)}. In short, we manually
modify a vote distribution by packing or cracking so as to flip a
single district from one party to the other. There are four possible
composite choices: whether we are packing or cracking and whether it
is the Republicans or the Democrats who are in charge of the
gerrymander. In reality, the votes from the flipped district could be
distributed among other districts by a combination of packing and
cracking, but we do not attempt to model this. We focus below on the
case in which the Republicans are flipping a single democratic
district to republican control using cracking. The other three cases
are treated similarly.

In practice, the details of a gerrymander will depend on many
factors. One such factor will be the geography of the state. If a
given district is being cracked so as to turn it from a democratic
district to a republican district, the surplus democratic voters will
have to be allocated to adjacent districts. Of course, this process
can be iterated by swapping other democratic voters for republican
voters in second-order neighbors of the original
district. Nonetheless, there are obvious geographic constraints that
may be significant. Another factor is how risk averse the
gerrymandering party is. For example, if the Democrats wish to
maximize their potential gain in seats (albeit at a high risk of the
plan backfiring) they can crack republican districts by creating
districts that are (say) 49\% republican. On the other hand, if the
Democrats feel the political winds will be against them in the
upcoming decade, they may prefer to pack Republicans into districts so
that the democratic districts are no more than (say) 35\% republican.

For our model, we make the following conventions for how the
gerrymander is achieved.
\begin{enumerate}
\item \emph{The flipped district is the democratic district that
  is won by the narrowest margin.}\label{ci}
\item \emph{The gerrymander does not create any new
  republican-majority districts with a democratic vote fraction of
  greater than $0.45$.} We choose this value on the basis that a 45--55
  split is frequently considered the threshold for a race to be
  competitive (see, for example,~\cite{abramowitz}). Any
  republican-majority district with a democratic vote fraction higher
  than this before the cracking is allowed to remain at such a
  level.\label{cii}
\item \emph{The modified democratic vote fraction in the flipped
  district is chosen according to a linear regression of the
  democratic vote fractions among the republican districts.} If the
  linear regression yields a new democratic vote fraction in the
  flipped district greater than $0.45$, then the value is set to
  exactly $0.45$.\label{ciii}
\item \emph{The democratic votes shifted from the flipped district
  are distributed evenly among the republican-majority districts
  with a democratic vote fraction of at most $0.45$.} In order to
  avoid violating the second convention, this process may need to be
  iterated (see Example~\ref{lab:example} below).\label{civ}
\end{enumerate}

In order to illustrate the method in practice we present the following
example of flipping a district from democratic to republican by
cracking.  While we use hypothetical data in this example, all
subsequent applications of simulated packing and cracking in this
paper involve vote distributions from actual elections.

\begin{example}\label{lab:example}
  Consider a $10$-district election 
  \begin{equation*}
    \bl = (0.37,0.40,0.43,0.46,0.60,0.63,0.66,0.69,0.72,0.75).
  \end{equation*}
  By Convention~\ref{ci}, we flip the fifth district. A linear
  regression of the four points
  \begin{equation*}
    \left\{\frac{i}{10}-\frac{1}{2\cdot 10},\ell_i\right\}_{i=1}^4 =
    \{(0.05,0.37), (0.15,0.40), (0.25,0.43), (0.35,0.46)\}
  \end{equation*}
  yields a line with intercept $0.355$ and slope $0.3$. Thus, according to the linear
  regression line, the flipped district should be switched from a
  democratic vote fraction of $0.6$ to one of $0.49$. However, by
  Convention~\ref{ciii} we instead choose a value of $0.45$. In order
  to maintain the same statewide democratic vote fraction, there must
  be a net increase of $0.15$ among the first three districts (note
  that the fourth district is not included since its democratic vote
  fraction already exceeds $0.45$). Convention~\ref{civ} instructs us
  to distribute these democratic votes evenly among the three
  districts. The resulting vote distribution is
  \begin{equation*}
    (0.42,0.45,0.48,0.46,0.45,0.63,0.66,0.69,0.72,0.75).
  \end{equation*}
  However, following Convention~\ref{civ}, we iterate the process by
  redistributing the excess fraction of $0.03=0.48-0.45$ from the
  third district evenly among the first two districts. Since the
  second district is already at a value of $0.45$, the amount is
  entirely distributed to the first district. This yields a final
  vote distribution of
  \begin{equation*}
    \bnl = (0.45,0.45,0.45,0.46,0.45,0.63,0.66,0.69,0.72,0.75).
  \end{equation*}
\end{example}

We utilize \spc\ for two purposes in this paper. In each instance, we
analyze a vote distribution using some function both before and after
the \spc. We have rerun the analyses in this paper with different
values of the maximum democratic vote fraction equal to $0.40$ and
$0.49$. We have also analyzed the effect of a ``greedy'' distribution
of votes in which the extra votes from the flipped district are
distributed to the districts whose democratic and republican votes are
distributed most unevenly. The resulting data and figures
corresponding to Figures~\ref{fig:pandcdec},~\ref{fig:cottrell-mpandc}
and~\ref{fig:olssim} are qualitatively very similar.

\section{The declination}
\label{sec:declination}

The declination function was introduced in~\cite{declination} as a way
of identifying potential gerrymanders by quantifying partisan
asymmetry in the vote distribution. A number of other
vote-distribution functions created for this purpose can be found in
the literature (several of these are mentioned in
Section~\ref{sec:recent}).

The declination is based on two observations. The first is that a
constitutionally manageable standard for partisan gerrymandering could
conceivably be based on some measure of partisan asymmetry, that is, a
measure of how the district plan treats the parties differently. The
second observation, which we make in~\cite{declination}, is that there
should be nothing special about the 50\% vote threshold in individual
districts. Combining these observations leads us to compute the ratio
of ``average winning margin'' to ``fraction of seats won'' for each
party. The declination is simply a comparison of these two ratios. If
the 50\% threshold is truly not special, then the resulting ratios for
each party should be approximately equal. The declination can be computed
geometrically as follows (see~\cite{declination} for details).

\begin{figure}
  \centering
  \includegraphics[width=.8\linewidth]{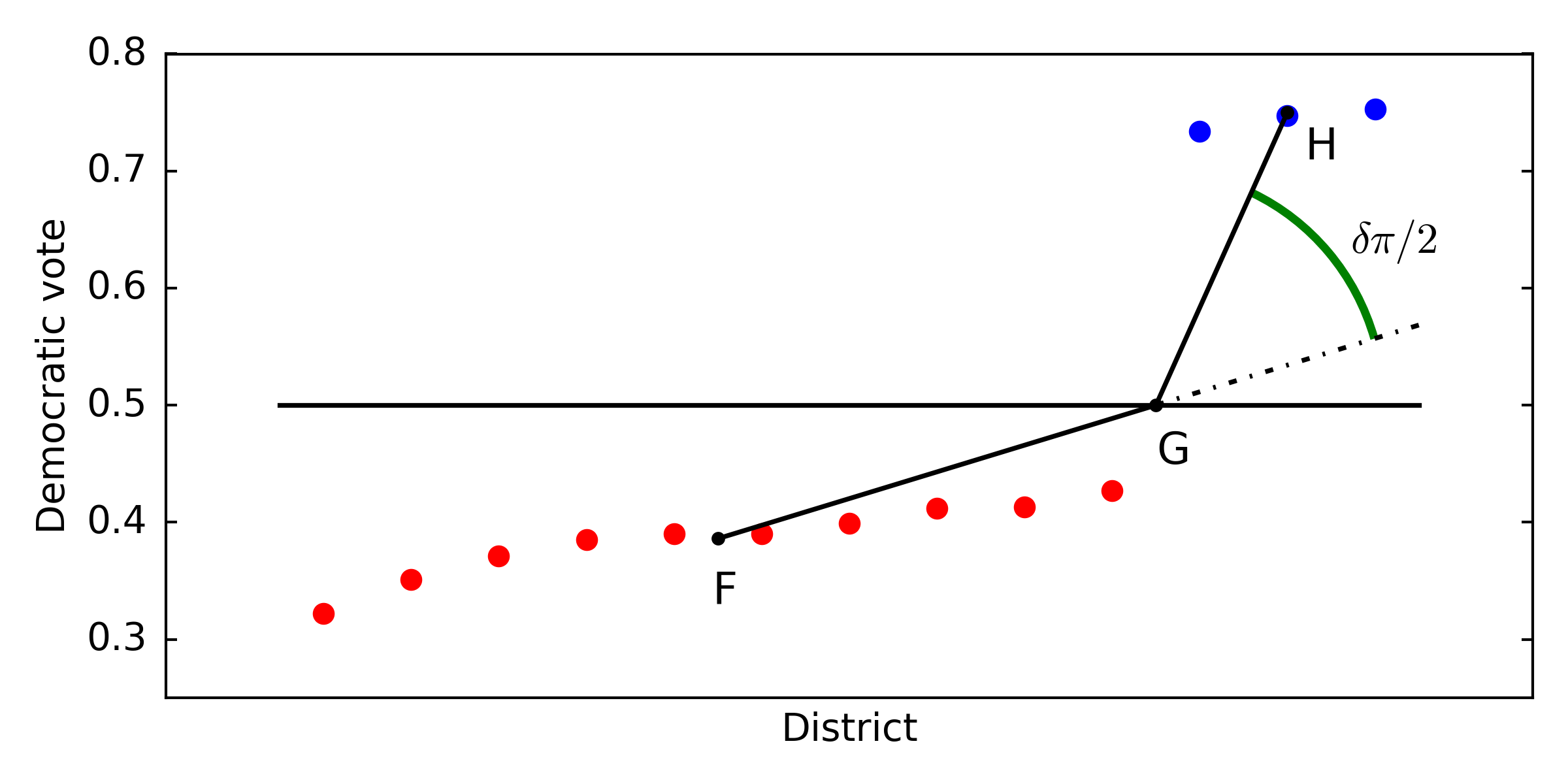}
  \caption{Illustration of the two lines $\overline{FG}$ and
    $\overline{GH}$ arising in the definition of the declination,
    $\delta$, for the $13$-district 2014 North Carolina congressional
    election. Districts have been sorted in increasing order of
    democratic vote share. The value of $\delta$ is approximately
    $0.54$ in this case.}
  \label{fig:angle}
\end{figure}

Let $v_i$ denote the point $(i/N-1/2N,\ell_i)$ for each $1\leq i\leq
N$. Place a point $F$ at the center of mass of the points $v_i$
corresponding to the districts the Democrats lose; a point $H$ at
the center of mass of the points $v_i$ the Democrats win; and a third
point $G$ at $(k/N,1/2)$, where $k$ is the number of districts the
Democrats lose. See Figure~\ref{fig:angle}.

If the district plan treats the parties symmetrically, we would expect
the point $G$ to lie on the line $\overline{FH}$. As such we define
the \emph{declination}, $\delta$, to be $2/\pi$ times the angle between the
lines $\overline{FG}$ and $\overline{GH}$ (using the convention that a
counterclockwise angle from $\overline{FG}$ to $\overline{GH}$ is
measured positively). Multiplication by $2/\pi$ converts from radians
to fractions of $90$ degrees. Therefore, possible values of the
declination are between $-1$ and $1$. Positive values indicate a
republican advantage while negative values indicate a democratic
advantage.

\subsection{The $S$-declination}

The declination metric was introduced for the purpose of measuring the
degree to which district plans result in partisan asymmetry in the
vote distribution. It is essentially an angle associated to each
election and it is not obvious that it can be used to provide a good
estimate of how many extra seats have been won due to the
asymmetry. While the declination itself does not count seats, we
propose that the \emph{$S$-declination}, which we set equal to
$5N\delta/12$, does.

\begin{figure}
  \centering
  \includegraphics[width=.9\linewidth]{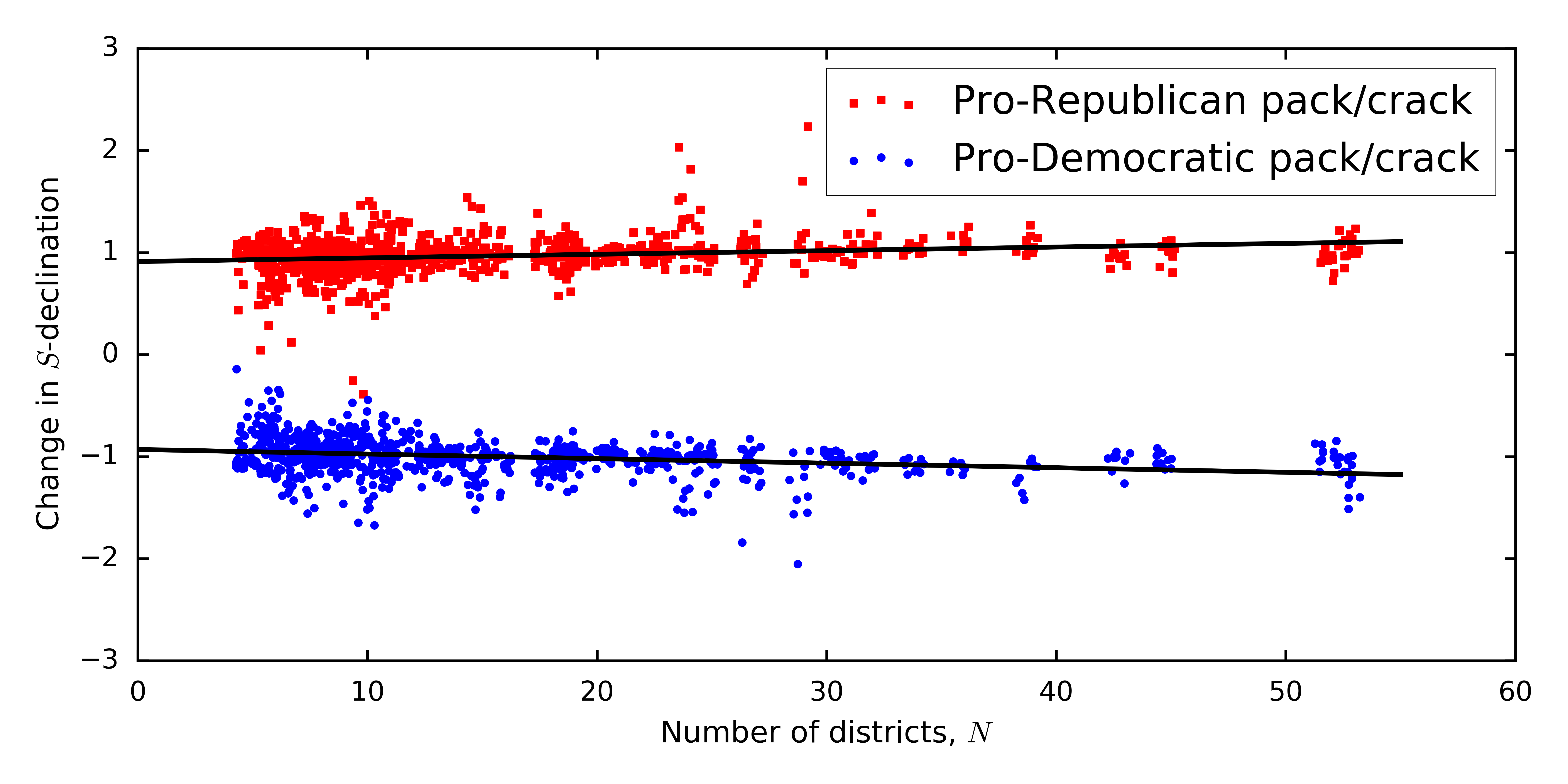}
  \caption{Plot of change in the $S$-declination due to \spc. Red
    squares indicate the 502 ways in which presidential-year House
    elections since 1972 can be packed/cracked so that the Republicans
    win one more seat, subject to the constraints noted in the main
    text; blue circles indicate the 491 corresponding ways in which
    presidential-year elections since 1972 can be packed/cracked so
    that the Democrats won one more seat. Note that horizontal
    coordinates have been jittered for clarity. The linear regression
    lines are $0.908 + 0.004 N$ ($r^2 = 0.05$, $\mathrm{RMSE}=0.17$)
    and $-0.9305 - 0.004 N$ ($r^2=0.09$, $\mathrm{RMSE} = 0.16$),
    respectively.}
  \label{fig:pandcdec}
\end{figure}

To support this claim we use \spc\ on the vote distributions from a
number of elections and examine the effect on the $S$-declination. If
the distribution is modified by flipping one district from democratic
to republican through packing or cracking, the $S$-declination should
increase by about one. If the flip is from republican to democratic,
it should decrease by about one. To this end, we consider the data set
from~\cite{declination} consisting of all state-wide elections to the
US House of Representatives in presidential-election years since
1972. (Note that a multilevel model is used in~\cite{declination} to
impute the democratic vote fraction for uncontested elections.) For
any such election we can attempt to pack or crack it in favor of
either party. In Figure~\ref{fig:pandcdec} we illustrate the change in
$S$-declination for those instances in which the \spc\ is successful.

The instances of \spc\ plotted in Figure~\ref{fig:pandcdec} are
restricted in two ways. First, we require that each party win at least
one seat both before and after the packing/cracking. This is necessary
for the $S$-declination of each distribution to be defined. Second, we
require that there be at least three districts into which to
distribute the votes from the flipped district. For example, when a
seat is being flipped from democratic to republican by cracking, we
require that there be at least three republican seats in the original
distribution. Together, these restrictions exclude states in which
there are four or fewer congressional districts. In the current
apportionment cycle there are 21 such states:
\begin{multline*}
  \text{AK, AR, DE, HI, IA, ID, KS, ME, MS, MT, ND,}\\ \text{NE,
  NH, NM, NV, RI, SD, UT, VT, WV, and WY.}
\end{multline*}
There were 352 state-year pairs in which there were at least five
districts and each party won at least one seat. Given the four
possible combinations of packing/cracking and
pro-republican/pro-democratic, this offers 1408 possible applications
of \spc. However, for 415 of these, either there was not enough room
for the chosen packing/cracking or one of the constraints was not
satisfied.

When flipping a seat from democratic to republican, we find that 95\%
of the time the $S$-declination changes by an amount between $0.69$
and $1.20$. For a flip from republican to democratic, the
corresponding range is $-1.26$ to $-0.74$. We conclude that the
declination is a reasonably good recorder of \pandc.

We estimate how many net seats have been won in each year on the basis
of partisan asymmetry in the vote distribution as follows: For a given
year add together the values of the $S$-declination (rounded to the
nearest integer) for each state. The results are shown in
Table~\ref{tab:net}. Unfortunately, the $S$-declination simply
provides an estimate without error bounds. Note that a given
state-year does not contribute to the sum from that year if one party
wins all of the seats in the state that year. However, we do not place
the additional constraints used for the generation of
Figure~\ref{fig:pandcdec} as the values in Table~\ref{tab:net} rely
only on actual vote distributions and not on values derived from \spc.

\begin{table}[ht]
  \centering
  \caption{Net effect of vote-distribution asymmetry on partisan
    composition of House as determined by $S$-declination. Positive
    (negative) values indicate a benefit to Republicans (Democrats).}
  \begin{tabular}{rcrcrcrcrc}\hline
    Year & Seats & Year & Seats & Year & Seats & Year & Seats & Year & Seats\\\hline
    1972 & -1   & 1982 & -16  & 1992 & -7   & 2002 &  5   & 2012 &  28  \\
    1974 & -2   & 1984 & -20  & 1994 &  0   & 2004 &  20  & 2014 &  20  \\
    1976 & -23  & 1986 & -10   & 1996 &  12  & 2006 &  20  & 2016 &  25  \\
    1978 & -28  & 1988 & -19  & 1998 &  9   & 2008 &  7   & \\
    1980 & -10   & 1990 & -12  & 2000 &  13  & 2010 &  17  & \\\hline
  \end{tabular}
  \label{tab:net}
\end{table}

\section{Two recent studies}
\label{sec:recent}

We now turn our attention to two recent studies that have also
attempted to answer the question of the net result of gerrymandering
in the House. 

\subsection{Royden and Li study}

The most recent study, a Brennan Center~\cite{brennan}
report by Royden and Li, considers House elections since 2012. The
authors' approach is to use functions (akin to the declination) that
identify gerrymanders by quantifying asymmetries in the vote
distribution among districts. They consider three such functions from
the literature: the efficiency gap~\cite{McGhee,M-S}, the seats-votes
curve~\cite{McGann} and the median-mean
difference~\cite{krasno,McDonaldBest,Wang}. The first two of these
functions are able to provide estimates of the number of seats won as
a result of gerrymandering. For the 2012 congressional elections, these
functions both come to very similar conclusions: the Republicans
gained between 25 and 37 extra seats. The estimate from
Table~\ref{tab:net} falls at the lower end of this range.
An important caveat to their methodology is that it does not account for any
inherent democratic disadvantage owing to geographic clustering. This
is true of the declination as well. However, the authors find a high
correlation between states with single-party redistricting control and
large numbers of extra seats, thereby suggesting that geography fails
to account for a significant portion of the asymmetry.

%

\subsection{Chen and Cottrell study}

In~\cite{cottrell}, Chen and Cottrell take what is perhaps a more
intuitive approach to evaluating the net effect of gerrymandering. For
each state, they use 200 computer-generated ``neutral'' district plans
as a standard to which the results of the enacted district plan are
compared. Utilizing aggregated precinct-level presidential vote data,
they estimate the probability that each simulated district in each
simulated plan elects a democratic representative. These
probabilities, over all districts, immediately yield an expected
number of democratic representatives in the House under each such
simulated district plan. They find that the expected number of
democratic seats in the enacted plans is only one less than the
average of the expected number of seats in the simulated plans,
leading them to conclude that the net effect of gerrymandering in the
House is trivial. Note that a putative advantage is that by
incorporating the geographic distribution of voters, this approach has
the potential to account for geographic clustering.

\begin{remark}
  Choosing simulated district plans from an appropriate distribution
  is a notoriously difficult problem that is not even well
  defined. Issues one must contend with include how to deal with
  constraints imposed by the Voting Rights Act; how to recognize
  communities of interest; the extent to which one should respect
  existing political boundaries; how to define compactness; how strict
  compactness constraints should be made; and how to weigh all of
  these disparate factors. Even once one makes such decisions, it is
  not necessarily clear how one chooses uniformly from the space of
  remaining ``acceptable'' district plans. Finally, there remains the
  nebulous problem of how to relate the choices made to what human map
  drawers do, or should be doing, when they create maps. We do not
  address these issues as they arise in~\cite{cottrell} as they are
  nuanced and beyond the scope of this paper.
\end{remark}

The declination discussed in Section~\ref{sec:declination} and the
functions utilized by Royden and Li directly count extra seats won as
a result of partisan asymmetry in the vote distribution. However
imperfect these measures may be, we have no more logically direct way
to count such seats. As mentioned above, the approach of Chen and
Cottrell has the advantage that it more directly addresses inherent
geographic advantages. However, while this approach should, \emph{a
  priori}, be able to count extra seats, the connection is less
direct. In the remainder of this section we argue, in fact, that the
logistic regression model used fails to capture gerrymandering to any
appreciable degree. In the computations that follow, we will omit
uncontested districts: Our goal is not to accurately estimate the
exact number of seats a given party would win, but simply to see how
the estimate changes under \spc.

We now present the notation required for the model. Let $p_i$ denote
the presidential vote in district $i$. Following Chen and Cottrell, we
consider a simple logistic-regression model for estimating the
expected number of democratic seats won in district $i$ as a function
of $p_i$.  According to their logistic model, the expected number of
democratic seats in district $i$ is $F(\beta_0+\beta_1 p_i)$ where
$F(x)=(1+e^{-x})^{-1}$ and $(\beta_0,\beta_1)$ are regression
coefficients that are estimated from the data. For an election with
presidential vote $\bp = (p_1,p_2,\ldots,p_N)$, the expected number of
democratic seats is then
\begin{equation}\label{eq:expd}
  E(\bp)=\sum_{i=1}^N F(\beta_0+\beta_1 p_i).
\end{equation}
Chen and Cottrell used presidential data from 2008 and 2012 along with
the four House elections from 2006 to 2012 to estimate the parameters
$\beta_0$ and $\beta_1$. In order to use election data from 1972 to
2012, we replaced $\beta_0$ and $\beta_1$ with random year effects
$\beta_{0[j]}$ and $\beta_{1[j]}$ with $j$ indexing the year. Under
this model (suppressing the dependence of $E(\bp)$ on $j$), the
expected number of democratic seats becomes
\begin{equation}\label{eq:expdmulti}
  E(\bp)=\sum_{i=1}^N F(\beta_{0[j]}+\beta_{1[j]} p_i).
\end{equation}
Estimates of the intercept and slope parameters $\beta_{0[j]}$ and
$\beta_{1[j]}$ using maximum likelihood are listed in
Table~\ref{tab:ols}.

Our approach to analyzing this model is analogous to our validation of
the $S$-declination: We apply \spc\ to historical vote distributions
and observe the effect on the predicted number of seats each party
will win. However, there is an added complication. When packing and
cracking, the vote of greatest interest in district $i$ is not the
presidential vote $p_i$, but the legislative vote
$\ell_i$. Unfortunately, \spc\ provides us only with a modified
legislative vote. We will need a mechanism for estimating
district-level presidential votes from district-level legislative
votes.

\begin{figure}
  \centering
  \includegraphics[width=\linewidth]{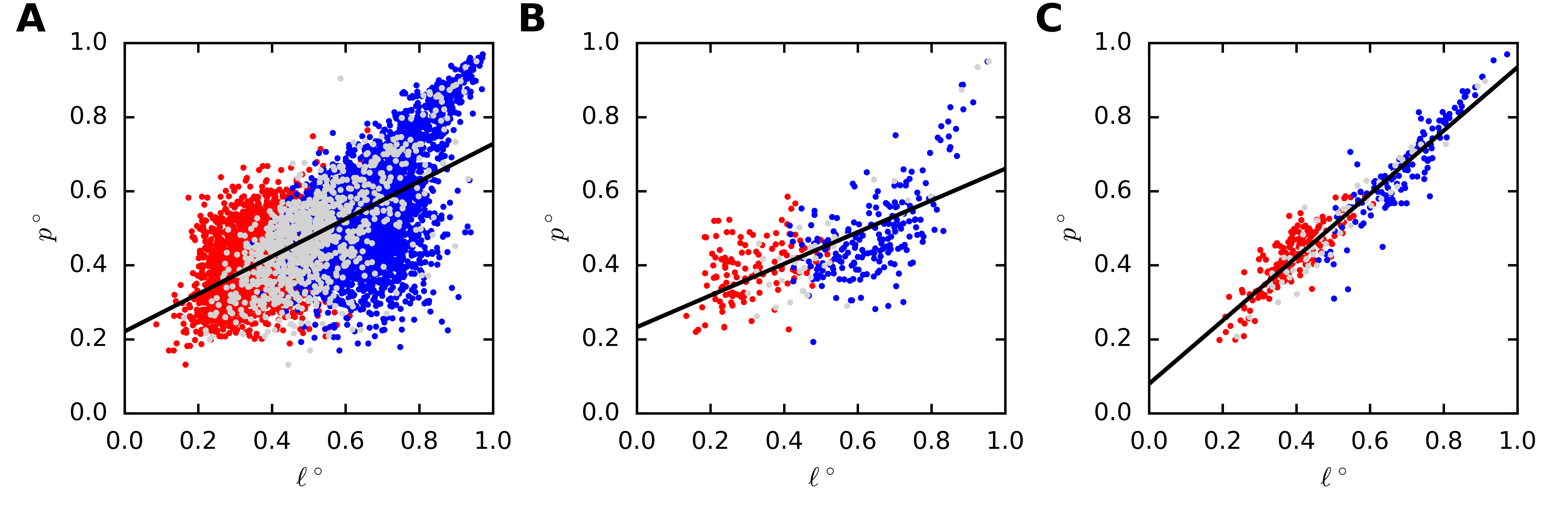}
  \caption{Plot of legislative vote $\bol$ versus presidential vote
    $\bop$. Races are colored according to incumbency statuses of the
    candidates: democratic incumbent (blue); republican incumbent
    (red); no incumbent or two incumbents from opposite parties
    (gray). Regression lines are also plotted. {\bf (A)}
    Presidential-year elections from 1972--2012: intercept $0.22$,
    slope $0.51$, $r^2 = 0.42$; {\bf (B)} 1980 election: intercept
    $0.23$, slope $0.43$, $r^2=0.44$; {\bf (C)} 2012 election:
    intercept $0.08$, slope $0.85$, $r^2=0.91$.}
  \label{fig:pvl}
\end{figure}

If $p_i$ and $\ell_i$ directly determined each other, there would be
no need for a logistic regression in~\eqref{eq:expd}. You would be
able to determine who won the seat from the value of $p_i$ without any
appeal to probability. As shown in Figure~\ref{fig:pvl} (see also
Table~\ref{tab:ols}), there is some justification for asserting a
linear relation between the presidential vote and the legislative
vote. (This relation is certainly much stronger in some years than
others.) One could also choose to fit the three subsets corresponding
to elections in which there is a democratic incumbent, a republican
incumbent, or no incumbent, however we do not pursue this variation.

In the remainder of this subsection we will model $p_i$ as a linear
function, $g$, of $\ell_i$. Letting $j$ index the year, we set $p_i =
g(\ell_i) := \gamma_{0[j]} + \gamma_{1[j]}\ell_i$ for some
coefficients $\gamma_{0[j]}$ and $\gamma_{1[j]}$. The function $g$ is
dependent on the year $j$, however we will suppress this from the
notation. For the remainder of this section we will reserve $\bop =
(\op_1,\ldots,\op_N)$ and $\bol = (\ol_1,\ldots,\ol_N)$ for the
historical values given to us in our data set. We reserve $\bnl =
(\nl_1,\ldots,\nl_N)$ for the legislative vote after applying one of
the four variations of \spc\ to $\bol$. The presidential vote
associated to $\nl_i$ will be written $\np_i = g(\nl_i)$. (Note,
however, that $g(\ol_i)$ does not equal $\op_i$.)

\subsubsection{The simplest model: $\gamma_{0[j]} = 0$, $\gamma_{1[j]}=1$ for all $j$.}
We begin with the simplest reasonable parameters: $\gamma_{0[j]} = 0$
and $\gamma_{1[j]} = 1$ for all $j$. While there are a number of
obvious objections to assuming the presidential and legislative votes
are equal, there are reasons to believe that it is a best-case
scenario for the Chen-Cottrell approach. That is, if their approach
does not record \spc\ under this model, then it will not do so under a
more tenuous relation between the two votes. Since $E(g(\bol))$ is the
expected number of democratic seats initially, $E(g(\bnl))-E(g(\bol))$
is the change in the expected number of democratic seats after
\spc. In Figure~\ref{fig:pandcdec}, a vertical coordinate of $1$
indicates one more seat for the Republicans. For consistency,
therefore, we consider the negative, $E(g(\bol)))-E(g(\bnl))$, the
expected change in the number of republican
seats. By~\eqref{eq:expdmulti} and our choice of $\gamma_{0[j]}=0$,
$\gamma_{1[j]}=1$, we then have that the expected change in republican
seats is given by
\begin{equation}\label{eq:zeroone}
  \begin{aligned}
  E(g(\bol))-E(g(\bnl))&= \sum_{i=1}^N [F(\beta_{0[j]}+\beta_{1[j]} g(\bol)_i) -
                                      F(\beta_{0[j]}+\beta_{1[j]} g(\bnl)_i)]\\
                       &= \sum_{i=1}^N [F(\beta_{0[j]}+\beta_{1[j]} (0 + 1\cdot \ol_i) -
                                      F(\beta_{0[j]}+\beta_{1[j]} (0 + 1\cdot \nl_i)]\\
                       &= \sum_{i=1}^N [F(\beta_{0[j]}+\beta_{1[j]} \ol_i) -
                                       F(\beta_{0[j]}+\beta_{1[j]} \nl_i))].
  \end{aligned}
\end{equation}
Note that we do not use $\bop$ in the above computation. No matter how
the function $g$ biases the answer, we at least want this bias to be
consistent both before and after the \spc.

\begin{figure}
  \centering
  \includegraphics[width=.8\linewidth]{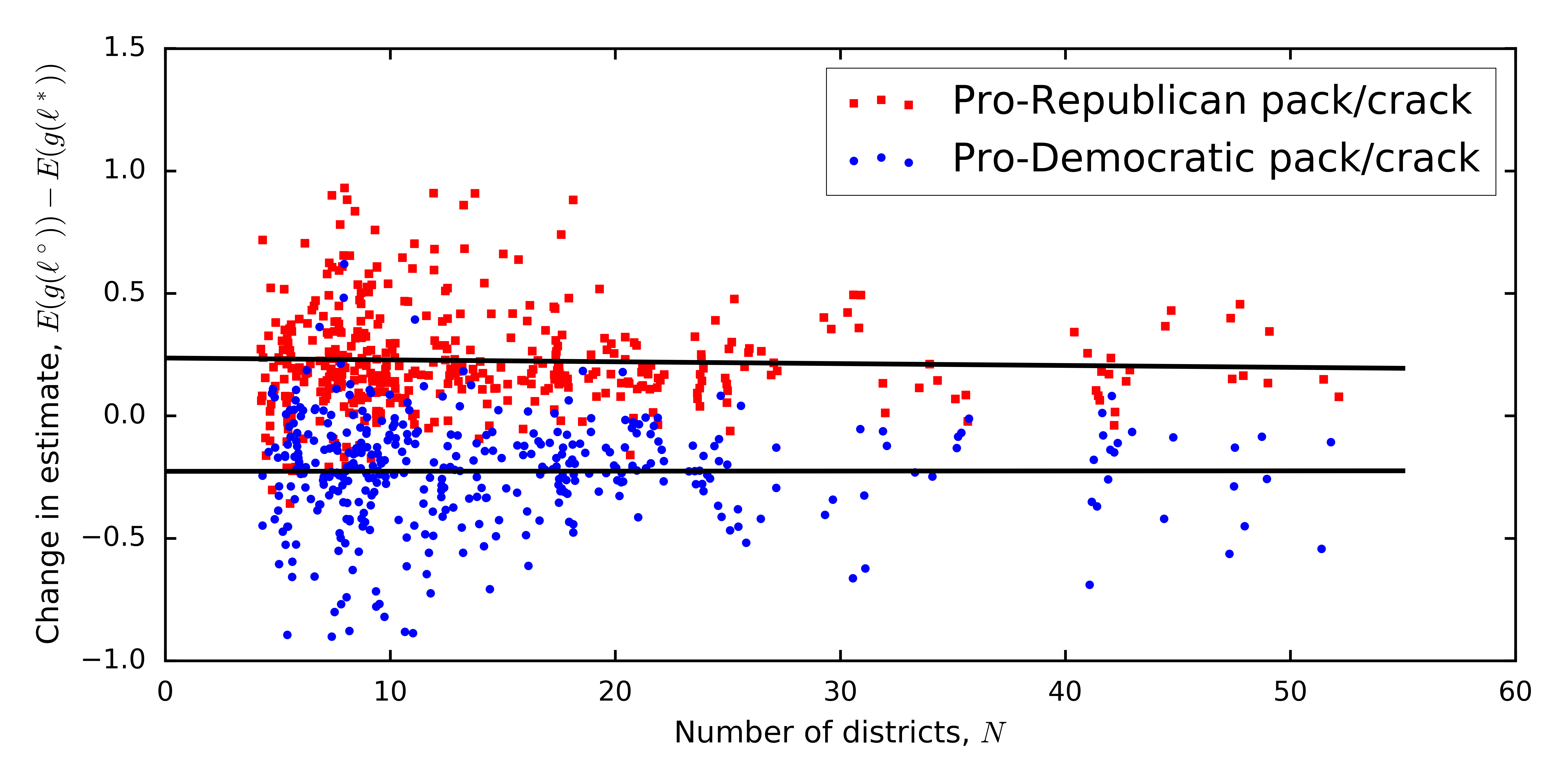}
  \caption{Change in the logistic-regression estimate of number of
    seats due to \spc\ using~\eqref{eq:zeroone}. Red squares correspond
    to the 418 ways in which House elections since 1972 can be
    packed/cracked so that the Republicans win one more seat, subject
    to the constraints noted in the main text; blue dots indicate the
    390 corresponding ways in which elections since 1972 can be
    packed/cracked so that the Democrats win one more seat. The linear
    regression lines are $0.236 - 0.001 N$ ($r^2 < 0.01$,
    $\mathrm{RMSE} = 0.207$) and $-0.227 + 0.000 N$ ($r^2 < 0.01$,
    $\mathrm{RMSE} = 0.216$), respectively.}
  \label{fig:cottrell-mpandc}
\end{figure}

Figure~\ref{fig:cottrell-mpandc} depicts values of the expression
in~\eqref{eq:zeroone} under the flipping of one seat via \spc. This
figure is analogous to Figure~\ref{fig:pandcdec}, which shows the
corresponding results for the $S$-declination.  If this model faithfully
recorded the effects of gerrymandering --- and hence packing and
cracking --- we would see the red squares vertically clustered around
$+1$ and the blue circles vertically clustered around $-1$. However,
as seen in Figure~\ref{fig:cottrell-mpandc}, this is not at all the case.


\subsubsection{Qualitative explanation for why Chen-Cottrell model fails.}

\begin{figure}
  \centering
  \includegraphics[width=.8\linewidth]{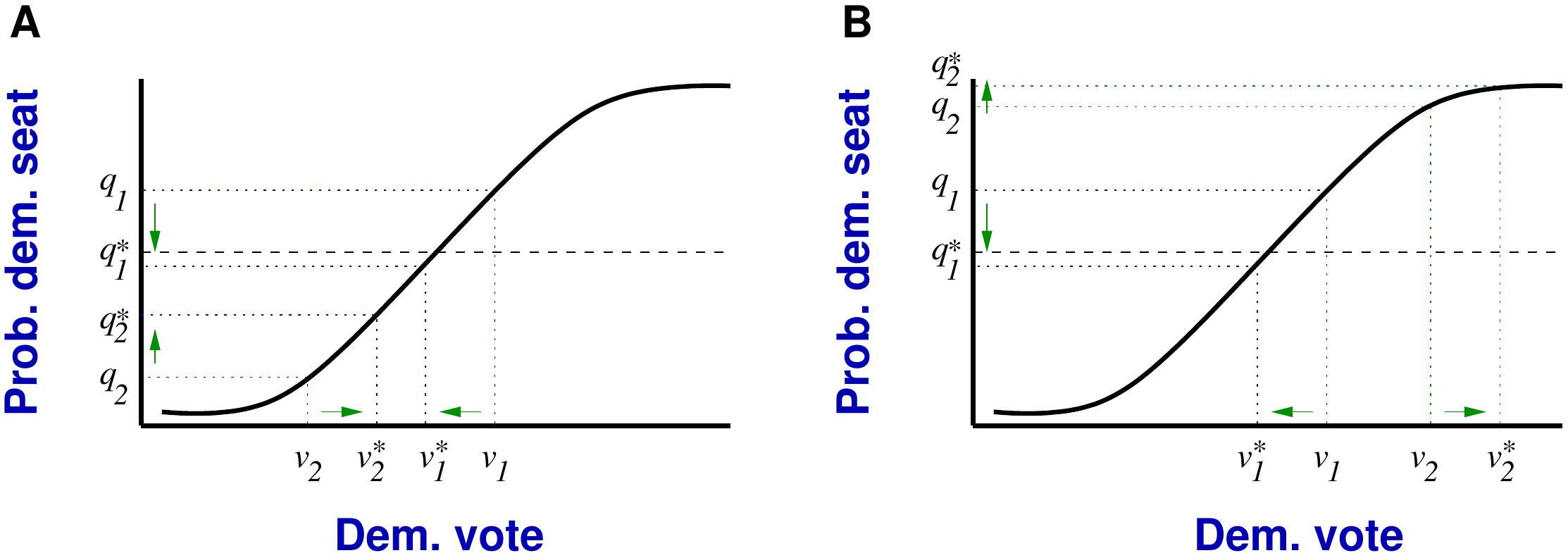}
  \caption{{\bf (A)} Example showing how cracking a district could
    lead to essentially no change in the expected number of democratic
    districts, as determined by the logistic model. {\bf (B)} Example
    showing how packing a district could lead to a moderate decrease
    in the expected number of democratic districts, as determined by
    the logistic model.}
  \label{fig:logistic-ex}
\end{figure}

We believe there is a simple reason for the relative invisibility of
packing and cracking to the logistic model. When districts are redrawn
so that, say a 55\% democratic district becomes 45\% democratic, the
map drawers are doing so with a detailed understanding of the partisan
composition of each affected district. While the gerrymander will
degrade over time and may be susceptible to wave elections, the
probability of the particular district electing a democratic
legislator goes from essentially one to essentially zero. However, the
probability of that district being democratic \emph{as determined by
  the fitted logistic curve} will change by a much smaller
amount. Furthermore, the decrease in probability for the given
district will be at least partially offset by increased probabilities
in the districts to which the democratic voters are now allocated.

Figure~\ref{fig:logistic-ex} illustrates how cracking or packing a
single district could lead to at most a small change in the expected
number of seats as determined by a sum over the probabilities obtained
from the logistic regression. Since we are setting $p_i = 0 + 1\cdot
\ell_i = \ell_i$, there is only a single vote fraction associate to
each district, which we denote by $v_i$. We write $q_i$ for $F(\beta_0
+ \beta_1 v_i)$, the probability of a democratic win in district
$i$. The corresponding values after \spc\ are denoted by $v^*_i$ and
$q^*_i$, respectively. Suppose District $1$ is being
flipped from republican to democratic and suppose District $2$ received
some of the reallocated democratic votes. In
Figure~\ref{fig:logistic-ex}.A, we see that the reduction in
probability from $q_1$ to $q^*_1$ is almost exactly offset by a
corresponding increase from $q_2$ to $q^*_2$. In
Figure~\ref{fig:logistic-ex}.B, there is almost no change from $q_2$
to $q^*_2$, leading to a net decrease in the expected number of
democratic seats. However, the net decrease is still much less than
one.

\subsubsection{Choosing $\gamma_{0[j]}$ and $\gamma_{1[j]}$ through linear regression.}

\begin{figure}
  \centering
  \includegraphics[width=.8\linewidth]{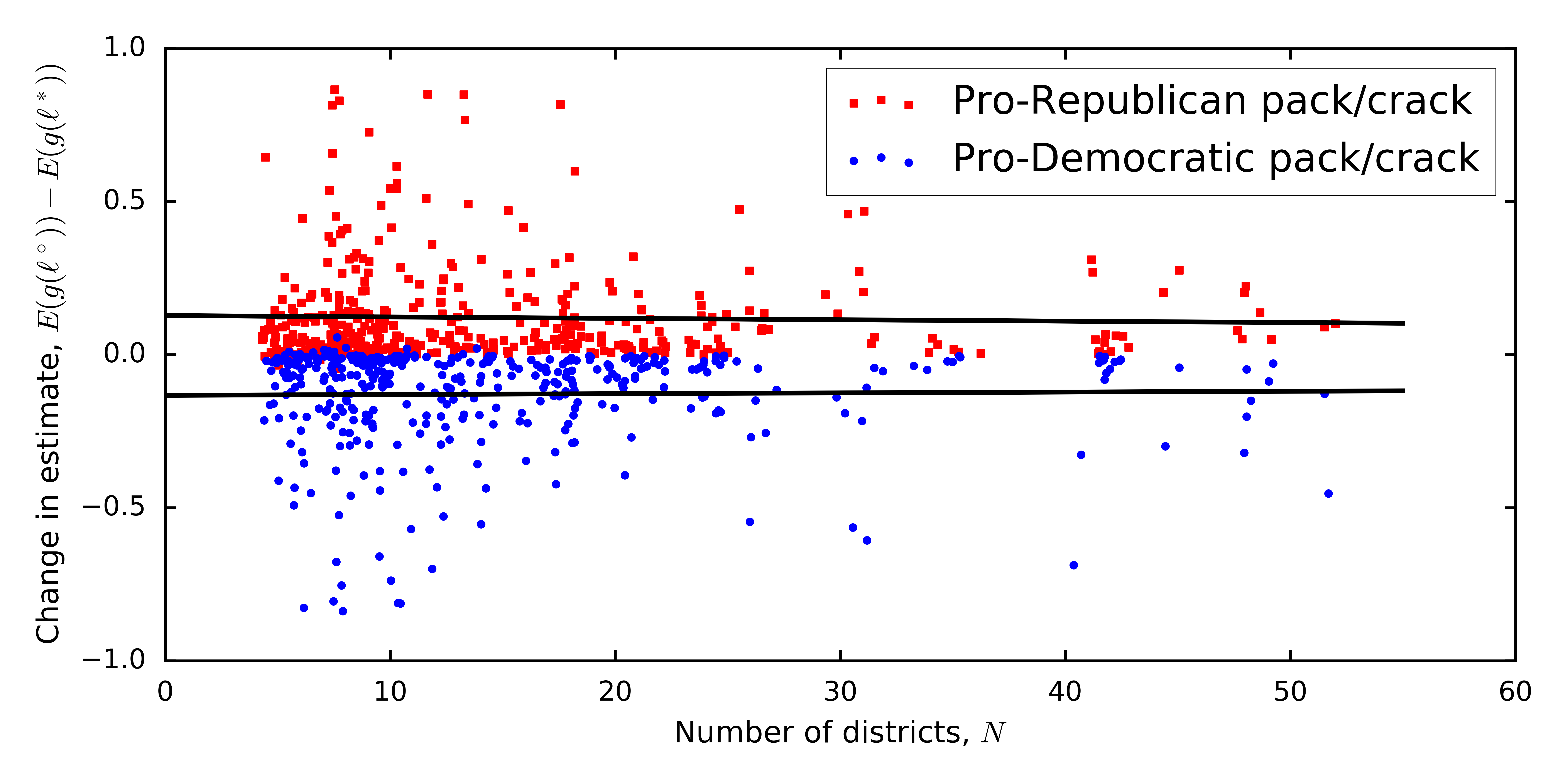}
  \caption{Plot of data in Figure~\ref{fig:cottrell-mpandc} using
    using~\eqref{eq:vard} instead of~\eqref{eq:zeroone}. The linear
    regression lines are $0.128 - 0.000 N$ ($r^2 < 0.01$,
    $\mathrm{RMSE} = 0.159$) and $-0.133 + 0.000 N$ ($r^2 < 0.01$,
    $\mathrm{RMSE} = 0.163$), respectively.}
  \label{fig:olssim}
\end{figure}

The choice of $(\gamma_{0[j]},\gamma_{1[j]}) = (0,1)$ for all $j$ is
useful for understanding the structural reasons for \emph{why} the
presidential-vote logistic model does not record partisan
gerrymandering. However, an ordinary linear regression fitted to the
presidential and legislative votes should yield a quantitatively
superior model. If we perform an ordinary linear regression using a
random effects model, we obtain the estimates of the coefficients
$\gamma_{0[j]}$ and $\gamma_{1[j]}$ displayed in
Table~\ref{tab:ols}. But these more defensible choices of
$\gamma_{0[j]}$ and $\gamma_{1[j]}$ do little to improve the
correspondence between \spc\ and the difference $E(g(\bol)) -
E(g(\bnl))$. In Figure~\ref{fig:olssim} we display the values of
\begin{multline}\label{eq:vard}
  E(g(\bol))-E(g(\bnl)) = \\\sum_{i=1}^N F(\beta_{0[j]} + \beta_{1[j]}(\gamma_{0[j]} + \gamma_{1[j]} \ol_i)) - 
  \sum_{i=1}^N F(\beta_{0[j]} + \beta_{1[j]} (\gamma_{0[j]} + \gamma_{1[j]} \nl_i))
\end{multline}
as $j$ indexes the years from 1972 to 2012. While there is better
separation between the two populations, the absolute change in the
expected number of seats is only about a tenth of a seat, even worse
than when we conflated legislative and presidential votes.

\begin{table}[ht]
  \centering
  \caption{Random effect estimates from linear and logistic
    models. The second and third columns contain estimates of the
    coefficients $\gamma_{0[j]}$ and $\gamma_{1[j]}$ for linear
    regression of presidential vote regressed on legislative vote for
    presidential election years using a random effects model where
    intercept and slope were random effects. The fourth and fifth
    columns list estimates for the intercepts and slopes for the
    random effects logistic regression of district outcome on
    presidential vote for presidential election years.}
  \begin{tabular}{ccccc}\toprule
    Year ($j$) & Intercept ($\gamma_{0[j]}$) & Slope ($\gamma_{1[j]}$) &
    Intercept ($\beta_{0[j]}$) & Slope ($\beta_{1[j]}$)\\\midrule
    1972 & 0.197 & 0.362 & -3.55  &  9.6\\
    1976 & 0.292 & 0.391 & -7.91  & 17.3\\
    1980 & 0.234 & 0.427 & -5.49  & 12.7\\
    1984 & 0.183 & 0.451 & -6.19  & 16.0\\
    1988 & 0.255 & 0.395 & -5.58  & 13.1\\
    1992 & 0.250 & 0.527 & -7.71  & 15.8\\
    1996 & 0.234 & 0.625 & -11.37 & 20.8\\
    2000 & 0.210 & 0.592 & -8.22  & 16.3\\
    2004 & 0.176 & 0.622 & -10.63 & 21.7\\
    2008 & 0.157 & 0.691 & -9.38  & 18.9\\
    2012 & 0.082 & 0.850 & -19.18 & 37.2\\\bottomrule
  \end{tabular}
  \label{tab:ols}
\end{table}

\subsubsection{Noise in the linear model.}\label{sec:noise}
One possibility worth examining is that the failure of the
Chen-Cottrell model in our simulations of \spc\ is due to our modeling
of the relationship between presidential vote and legislative vote as
linear. The linear relationship is much truer in some years than
others and, with the exception of 2012, is never completely
convincing. But arguing against this possibility is the fact that data
points from 2012 are included in Figures~\ref{fig:cottrell-mpandc}
and~\ref{fig:olssim} and there is no subpopulation clustered around
$\pm 1$. In fact, the sensitivity \emph{does} seem better for 2012
with a median (absolute-value) change of just over one
half. Regardless, any ``noise'' added to the relationship will merely
impede the ability of the presidential-vote logistic regression model
to record packing and cracking.

\begin{figure}
  \centering
  \includegraphics[width=.8\linewidth]{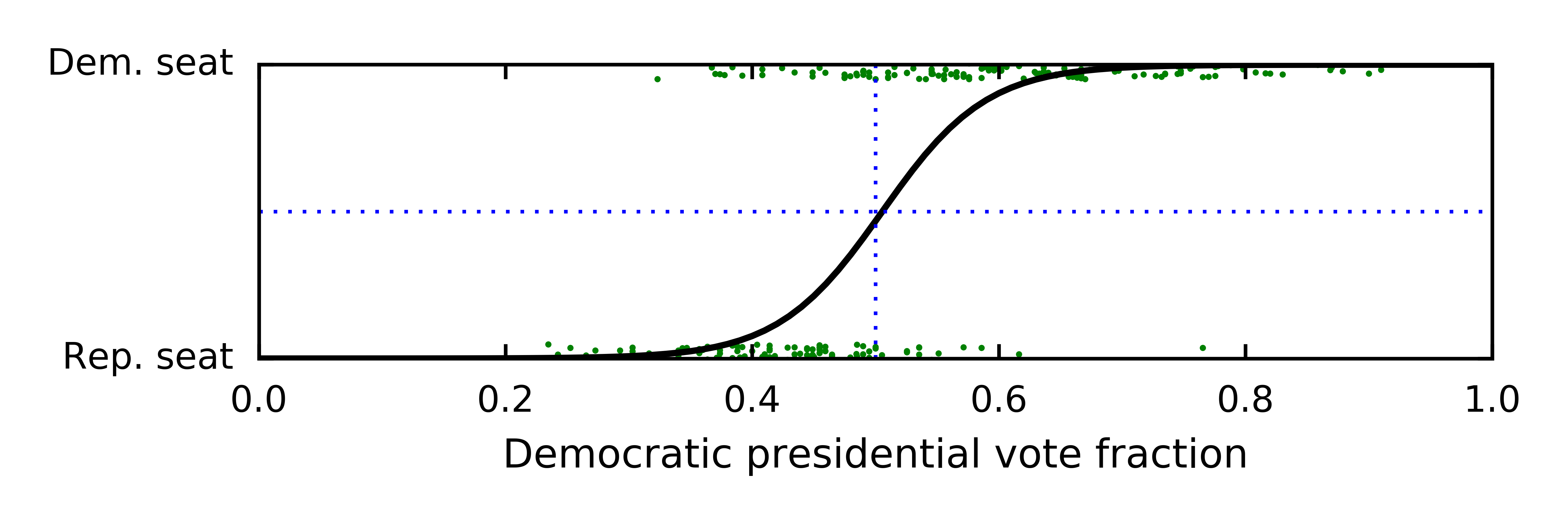}
  \caption{Plot of democratic presidential vote fraction versus
    legislative seat winner (points have been jittered) for the year
    2008. The fitted logistic curve with parameters $\beta_0 = -9.38$
    and $\beta_1 = 18.9$ (from Table~\ref{tab:ols}) is also shown.}
  \label{fig:logex}
\end{figure}

A model that postulates a linear relation between the proportion of
legislative and presidential vote in a district induces a probability
model for the probability of a democratic win in a district given the
presidential vote.  It can be shown that the effect of presidential
vote in the probability model decreases to zero as the correlation
between legislative and presidential vote decreases to zero.  In other
words, if there is a significant amount of noise in the relation
between legislative and presidential vote, presidential vote will be a
weak predictor of the winner of a district, further degrading the
ability of the logistic model to capture the effects of
gerrymandering.

\begin{remark}
Incumbent US House Representatives typically win upwards of 90\% of
their races. In particular, the fitted logistic regression curves will
be very different for the classes of 1) democratic incumbents, 2)
republican incumbents and 3) no incumbents. There is no reason
incumbency cannot be incorporated into the Chen-Cottrell model, but
there is one very important issue: The assumptions made for the
distribution of incumbents will have a significant impact on the
conclusions. Any historical gerrymander will change the relative
proportion of incumbents which, in turn, will affect the future
probabilities for who wins which seats.
\end{remark}




\section{Conclusion}
\label{sec:conc}

If the net impact of partisan gerrymandering on the US democracy is
understood to be minimal on a national level, there are likely to be
only piecemeal efforts to mitigate its influence. If, on the other
hand, its effect is shown to be large, there is likely to be a greater
political and judicial will to take steps to counter it on a national
level. Unfortunately, as detailed in~\cite{cottrell}, there are
manifest difficulties in directly measuring the net effect of
gerrymandering. The Chen-Cottrell approach inarguably addresses some
of these difficulties. For example, by using contemporaneous electoral
data, it can remove year-to-year effects. And, assuming the
simulations draw from the space of all districts in an appropriate
manner, it can account for geographic clustering. (Unfortunately,
there is no objectively ``correct'' distribution to draw from. Even
showing that the draws are sufficiently random is a difficult matter.)

However, as we have attempted to show above, performing a logistic
regression on presidential data is fatally flawed for this particular
purpose. Not only do our simulations indicate that it does not
effectively record packing and cracking, but we believe there are
convincing theoretical reasons for concluding it does not.  The
$S$-declination, defined and studied in this paper, provides a measure
of the number of seats gained/lost through partisan gerrymandering.
The validity of the $S$-declination for this purpose is strongly
supported by simulated packing and cracking
(Figure~\ref{fig:pandcdec}).

Neither we nor Royden and Li take into account geographic clustering
that has been shown to exist in~\cite{chen}. Nonetheless, their
results and our Table~\ref{tab:net} suggest that the net number of
seats won by gerrymandering is likely to be significant. And, as noted
in~\cite{declination}, it is not clear to what degree district plans
should be allowed to exacerbate (or mitigate) inherent geographic
distributions. Regardless, if there is only a minimal net effect of
gerrymandering on a national level, we find no evidence
in~\cite{cottrell} to support this position.

\section{Acknowledgments}

The US congressional data through 2014 was provided
by~\cite{jacobson}. The election data was analyzed using the
python-based SageMath~\cite{sage}. See~\cite{declination} for packages
used for, and details of, the imputation of votes. Additional Python
packages employed in this paper were Matplotlib~\cite{matplotlib} for
plotting and visualization and SciPy~\cite{scipy} for statistical
methods. Estimated coefficients for the linear and logistic
regressions were computed using the lme4 package~\cite{lme4} in
R~\cite{R}. We thank Jordan Ellenberg for suggesting a second
\spc\ algorithm to check.

\bibliography{gerrymandering}

\bibliographystyle{alpha}

\end{document}